\documentclass[prb,floatfix,twocolumn,showpacs,amsmath,amssymb]{revtex4}
\usepackage{epstopdf}
\usepackage{epsfig}
\usepackage{dcolumn}
\usepackage{bm}
\usepackage{color}

\usepackage[normalem]{ulem}

\newcommand{\dagga}{{\phantom{\dagger}}}

\begin{document}

\title{Spontaneous symmetry breaking in correlated wave functions}
\author{Ryui Kaneko,$^1$ Luca F. Tocchio,$^{2}$ Roser Valent\'\i,$^{1}$
Federico Becca,$^{2}$ and Claudius Gros$^{1}$}
\affiliation{
$^{1}$Institute for Theoretical Physics, University of Frankfurt, 
       Max-von-Laue-Stra{\ss}e 1, D-60438 Frankfurt a.M., Germany \\
$^{2}$Democritos National Simulation Center, Istituto Officina dei 
       Materiali del CNR, and SISSA-International School for Advanced 
       Studies, Via Bonomea 265, I-34136 Trieste, Italy
            }

\date{\today} 

\begin{abstract}
We show that Jastrow-Slater wave functions, in which a density-density 
Jastrow factor is applied onto an uncorrelated fermionic state, may possess
long-range order even when all symmetries are preserved in the wave function. 
This fact is mainly related to the presence of a sufficiently strong Jastrow 
term (also including the case of full Gutzwiller projection, suitable for 
describing spin models). Selected examples are reported, including the 
spawning of N\'eel order and dimerization in spin systems, and the stabilization 
of charge and orbital order in itinerant electronic systems. 
\end{abstract}

\pacs{71.27.+a, 71.10.Fd, 75.10.-b, 75.10.Jm}

\maketitle

\section{Introduction}\label{sec:intro}

Exact ground-state wave functions are known only for a limited number of 
many-body Hamiltonians (with exact solutions for the entire spectrum being 
even rarer)~\cite{sutherlandbook}. Variational states provide, alternatively, 
educated guesses for the ground state and for low-energy excitations. As they are
not related to particular weak-coupling approximations, variational approaches 
allow one to investigate nonperturbative effects. Nevertheless, they 
rely on an initial guess and may therefore sometimes be biased. Well-known 
examples of variational states are given by the Bardeen-Cooper-Schrieffer 
(BCS)~\cite{bcs1957} and Laughlin~\cite{laughlin1983} wave functions, 
describing, respectively, conventional superconductivity  and the fractional 
quantum Hall effect. Variational states have also been widely used ever since
Gutzwiller's seminal work on the Hubbard model~\cite{gutzwiller1963} in the 
context of correlated electronic and bosonic systems. A few benchmarking 
studies with other available many-body computational methods have been 
performed recently in the framework of the fermionic Hubbard 
model~\cite{tocchio2013,leblanc2015}.

Within the variational approach, it is easy to describe quantum phase
transitions. Usually, this is achieved by considering Hartree-Fock states,
which contain a suitable order parameter, whose finite value indicates the
stabilization of a symmetry-broken phase. One simple example is given by the 
half-filled Hubbard model on the honeycomb lattice, where antiferromagnetic 
order develops when the ratio between the on-site Coulomb interaction $U$ and 
the nearest-neighbor hopping $t$ exceeds a critical value~\cite{sorella1992}.
Most importantly, within the Hartree-Fock approach, the presence of long-range
order is obtained from an initial guess of the ordered pattern that is 
included into the wave function, thus implying an {\it explicit} symmetry
breaking.

In this paper, we want to assess the possibility that symmetry-broken phases 
can be obtained by using {\it symmetry-invariant} wave functions, which implies
that long-range order is obtained as a true spontaneous symmetry-breaking 
phenomenon. Even though no {\it explicit} bias is included in the wave functions,
one must keep in mind that this approach cannot provide a completely unbiased way
of obtaining {\it any} possible pattern for spin and/or charge order. We will 
show examples in which relatively simple orders emerge in symmetric states, 
while it remains a very hard task to devise a scheme in which a given wave 
function may describe many different spin and/or charge patterns that can be 
selected by tuning a few (variational) parameters.

In the context of spontaneous symmetry breaking, a well-known example is given 
by the Liang, Doucot, and Anderson (LDA) wave function, which was proposed to 
investigate quantum magnetism in the Heisenberg model on a square 
lattice~\cite{liang1988}. The LDA state is written in terms of {\it bosonic} 
degrees of freedom (e.g., singlets that cover the entire lattice), and it
embodies a possible representation of the resonating-valence-bond (RVB) 
states~\cite{anderson1987}. The LDA wave function is fully characterized by the 
weight factor $h(r)$ for a singlet of length $r$. To evaluate any 
expectation value over the LDA wave function, one must devise a stochastic 
sampling, based upon the Monte Carlo technique. Indeed, given the exponential 
increase of the dimension of Hilbert space, an exact treatment can be 
afforded only on very small clusters. Even though the LDA wave function does 
not break spin and lattice symmetries, it may describe magnetically ordered 
phases. This is the case when $h(r)$ decays slowly with $r$ [e.g., 
$h(r) \propto 1/r^p$, with $p<3.4$ on the square lattice]; by contrast, if 
$h(r)$ decays rapidly with $r$ (e.g., $p>3.4$), then the LDA wave function is 
magnetically disordered~\cite{liang1988,beach2007}. The great limitation of 
this wave function is that efficient Monte Carlo calculations can be afforded 
only in the presence of the Marshall sign rule~\cite{marshall1955}. In the absence 
of this rule, such as for triangular and kagome lattices, calculations suffer 
from a severe sign problem and only small cluster sizes can be 
afforded~\cite{sindzingre1994}. Therefore, its properties are well established 
in only a few cases.

Here, we consider an alternative approach and assess the possibility of having 
spontaneous symmetry breaking in a different family of quantum states, which 
are constructed from {\it fermionic} degrees of freedom, suitable to describe 
both itinerant (i.e., Hubbard) and localized (i.e., Heisenberg) systems. 
In the latter case, these fermionic wave functions give rise to alternative 
representations of RVB states~\cite{anderson1987}. Moreover, dealing with 
fermions has the notable advantage that Monte Carlo calculations can be easily 
performed on large clusters and any lattice geometry. The simplest of this class of 
variational states is the well-known Gutzwiller wave function that was 
introduced to deal with correlated electron systems~\cite{gutzwiller1963}:
\begin{equation}\label{eq:gutzslater}
|\Psi_g \rangle = {\cal P}_g |\Phi\rangle,
\end{equation}
where $|\Phi\rangle$ is a noninteracting fermionic state that is obtained
by filling $N$ given orbitals labeled by some index $\gamma$ (so that 
$|\Phi\rangle$ is an $N$-electron state):
\begin{equation}\label{eq:slater}
|\Phi\rangle = \prod_{\gamma=1}^{N} \phi^\dag_{\gamma} |0\rangle.
\end{equation}
In practice, $|\Phi\rangle$ can be obtained as an eigenstate (usually the
ground state) of a noninteracting Hamiltonian, containing, for example, hopping
and pairing terms (in the presence of pairing between up and down electrons, one 
can always perform a particle-hole transformation to have a Hamiltonian that 
commutes with the particle number, thus defining ``orbitals''). Finally, 
${\cal P}_g$ is the so-called Gutzwiller factor, which depends upon the 
variational parameter $g$:
\begin{equation}\label{eq:softgutz}
{\cal P}_g = \exp \left( -g \sum_{i} n_{i,\uparrow} n_{i,\downarrow} \right),
\end{equation}
where $n_{i,\sigma}$ is the electron density per spin $\sigma$ on the site $i$.
The role of this term is to reduce the amplitudes of electron configurations 
with doubly occupied sites, thus tuning the level of electron correlation:
$g=0$ corresponds to noninteracting particles, while $g=\infty$ totally
projects out configurations with doubly occupied sites, hence corresponding to
the strongest possible electron-electron interaction. We would like to remind
the reader that, also for fermionic wave functions such as the one given in
Eq.~(\ref{eq:gutzslater}), a Monte Carlo sampling is necessary to evaluate
any expectation value for large system sizes. Indeed, in the presence of
any correlation term, such as the Gutzwiller factor, analytical calculations are
not possible in lattices of generic dimensionality.

In the following, $N$ and $L$ denote the number of electrons and lattice 
sites, respectively; $n=N/L$ is the electron density. By restricting to fully
symmetric $|\Phi\rangle$, the correlated wave function~(\ref{eq:gutzslater}) 
may describe metallic or superconducting phases for generic densities $n$; 
insulating phases are possible only at half-filling $n=1$ and in the presence 
of a full Gutzwiller projector $g=\infty$~\cite{yokoyama1987a,yokoyama1987b}:
\begin{equation}\label{eq:fullgutz}
{\cal P}_\infty = \prod_i \left( 1 - n_{i,\uparrow} n_{i,\downarrow} \right).
\end{equation}
The is because the Gutzwiller term only correlates 
electrons on the same site: once charge excitations (holon-doublon couples) 
are created, the holon and the doublon are free to move around without any 
further penalization, thus leading to nonzero conductivity. 

When $g=\infty$ and $n=1$, charge degrees of freedom are completely frozen 
(i.e., there is exactly one electron on each site) and an insulator is obtained.
Nevertheless, the fully projected state:
\begin{equation}\label{eq:gutzwiller}
|\Psi\rangle = {\cal P}_\infty |\Phi\rangle
\end{equation}
still contains nontrivial spin degrees of freedom, so that it can be used
to study Heisenberg models~\cite{gros1989}.

A generalization of the Gutzwiller wave function~(\ref{eq:gutzslater}) can be
obtained by including density-density correlations at different sites, and it is 
given by the Jastrow-Slater state:
\begin{equation}\label{eq:jastrowslater}
|\Psi_J \rangle = {\cal J} |\Phi\rangle,
\end{equation}
where the Jastrow term includes correlations on different sites:
\begin{equation}\label{eq:jastrow}
{\cal J} = \exp \left( -\frac{1}{2} \sum_{i,j} v_{i,j} n_i n_j \right);
\end{equation}
here $v_{i,j}$ is a pseudo-potential for density fluctuations (the on-site 
term $v_{i,i}$ corresponds to the Gutzwiller parameter $g$) and
$n_i = \sum_{\sigma} n_{i,\sigma}$ is the total density on site $i$.
While long-range density-density correlations are crucial to describe a pure 
Mott insulator~\cite{capello2005}, here we will consider very simple Jastrow 
factors including only on-site and nearest-neighbor terms. In fact, already 
with this simple form, it is possible to describe situations in which 
symmetry-broken phases appear. Of course, long-range terms would be necessary 
also when considering more complicated charge/spin patterns. 

The generalization to multi-orbital models is also straightforward: one should
add orbital degrees of freedom in the noninteracting state $|\Phi\rangle$ 
(i.e., consider a noninteracting Hamiltonian with more than one orbital per
site) and introduce a Jastrow factor that couples density fluctuations on 
different sites and orbitals:
\begin{equation}\label{eq:jastrowmo}
{\cal J} = \exp \left( -\frac{1}{2} \sum_{i,j,\alpha,\beta} 
v^{\alpha,\beta}_{i,j} n^\alpha_i n^\beta_j \right),
\end{equation}
where $v^{\alpha,\beta}_{i,j}$ is a pseudo-potential for density fluctuations 
($v^{\alpha,\alpha}_{i,i}=g$, while $v^{\alpha,\beta}_{i,i}$ with 
$\alpha \ne \beta$ is the inter-orbital Gutzwiller parameter) and 
$n^\alpha_i = \sum_{\sigma} n^\alpha_{i,\sigma}$ is the charge density on the 
orbital $\alpha$ at  site $i$.

We will show that different kinds of spontaneous symmetry-breaking phenomena 
are possible within Jastrow-Slater wave functions, i.e., when using 
Eq.~(\ref{eq:jastrowslater}): more precisely, even when both the noninteracting 
state $|\Phi\rangle$ and the Jastrow factor ${\cal J}$ preserve all the lattice 
and spin symmetries, clear signatures of order can be obtained. For example, in 
the case of a discrete symmetry breaking, e.g., charge order, clear evidence 
of ergodicity breaking is detected when using single-particle moves in the Monte 
Carlo calculations. The use of fully symmetric wave functions allows us to describe 
quantum phase transitions by varying one parameter inside the variational wave 
function; for example, charge-density order is obtained in a system of itinerant 
electrons for a sufficiently strong nearest-neighbor Jastrow pseudo-potential 
(e.g., the one-dimensional lattice with $n=1/2$ filling and the triangular 
lattice with $n=2/3$ filling). These results can be understood thanks to a 
simple mapping from quantum averages to a {\it classical} problem of interacting 
particles. Then, the presence of a quantum phase transition when changing the 
variational state is directly connected to the existence of a classical phase 
transition in the related classical model.

In addition, we will also report the presence of antiferromagnetic long-range 
order in spin models, i.e., when using Eq.~(\ref{eq:gutzwiller}), similarly to 
what has been shown by using the LDA wave function. In this case, although the 
magnetization is exactly zero for all finite sizes (the quantum state is a spin 
singlet), magnetic order can be obtained in two dimensions whenever a suitable 
parametrization is considered. Furthermore, we will show that, within this 
class of fermionic states, the correct behavior is obtained in one dimension,
namely spontaneous breaking of SU(2) symmetry does not occur, in agreement 
with the Mermin-Wagner theorem~\cite{pitaevskii1991} and in contrast to bosonic
states~\cite{lin2012}. Instead, in one dimension, fermionic states may describe 
both gapless and dimerized (gapped) states, in agreement with the 
Lieb-Schultz-Mattis theorem~\cite{lieb1961}.

The paper is organized as follows: In Sec.~\ref{sec:density_order}, we show
the results for the appearance of charge order for itinerant electrons in one 
spatial dimension and in the triangular lattice, as well as the emergence of 
orbital order in a two-band model on the square lattice; we will also see that 
the emergence of charge order can be understood by mapping the wave function 
into the classical counterpart. In Sec.~\ref{sec:magnetic_order}, we present 
the results for magnetization and dimerization by applying the fully projected 
wave function where the classical mapping is no longer available; we first show 
that the wave functions reproduce the correct behaviors in one-dimensional spin 
models, and then we examine how magnetic order appears in two-dimensional spin 
models. Finally, in Sec.~\ref{sec:conclusions} we draw our conclusions.
 
\section{Charge-density and orbital order}\label{sec:density_order}

\subsection{The classical mapping}

Certainly, charge order can be obtained when using a Jastrow factor or a 
Slater determinant that breaks translational 
invariance~\cite{watanabe2005,miyazaki2009,tocchio2014}. However, this is an 
expected outcome, which will not be treated here; instead, as discussed above, 
we are interested in the more subtle case in which charge (or orbital) order 
may be settled in a perfectly symmetry-invariant variational state.

The variational calculation with the wave function~(\ref{eq:jastrowslater}) 
can be shown to correspond to a classical problem at finite 
temperature~\cite{motrunich2004,capello2006,capello2008}. This correspondence 
is very useful for showing that quantum phase transitions are possible within 
this class of variational states. To prove the mapping, let us consider a basis 
set $|x\rangle$ in which particles have definite positions in the lattice. 
For all operators $\theta$ that are diagonal in this basis, the quantum average:
\begin{equation}
\langle \theta \rangle = \frac{\langle \Psi | \theta |\Psi \rangle}
{\langle \Psi|\Psi \rangle}
\end{equation}
can be written in terms of the {\it classical} distribution:
\begin{equation}
\langle \theta \rangle = \sum_x P(x) \langle x|\theta|x \rangle,
\end{equation}
where $P(x)$ is given by:
\begin{equation}
P(x)=\frac{|\langle x|\Psi \rangle|^2}{\langle \Psi|\Psi \rangle}.
\end{equation}
Since $P(x) \ge 0$, there is a precise correspondence between the wave function
and an effective classical potential $V_{\rm cl}(x)$:
\begin{equation}\label{eq:classical}
P(x) \equiv \frac{1}{\cal Z} e^{-\beta_{\rm cl} V_{\rm cl}(x)},
\end{equation}
where $T_{\rm cl}=1/\beta_{\rm cl}$ represents an effective classical 
temperature. The explicit form of the potential $V_{\rm cl}(x)$ depends upon 
the choice of the Jastrow factor and the form of the noninteracting state 
$|\Phi \rangle$:
\begin{equation}\label{eq:classicalpot}
\beta_{\rm cl} V_{\rm cl}(x) = \sum_{i,j} v_{i,j} n_i(x) n_j(x)
                    - 2 \ln {\rm det} \Phi(x), 
\end{equation}
where $n_i(x)$ is the electron density at site $i$ for the configuration 
$|x\rangle$, i.e., $n_i|x\rangle = n_i(x)|x\rangle$ and 
$\Phi(x)=\langle x|\Phi\rangle$ is the amplitude of the noninteracting 
state over the configuration $|x\rangle$~\cite{motrunich2004,capello2006}.
The first term of Eq.~(\ref{eq:classicalpot}) is a two-body potential, which 
describes a classical model of oppositely charged particles (holons and 
doublons) mutually interacting through a given potential. In the presence of 
the second term in Eq.~(\ref{eq:classicalpot}), $V_{\rm cl}(x)$ is no longer 
a two-body potential. However, when density fluctuations are suppressed (by the 
Gutzwiller factor), the quadratic term gives the most relevant contribution, 
hence the mapping onto a classical model of interacting particles still holds with 
$\beta_{\rm cl} V_{\rm cl}(x) \simeq \sum_{i,j} v_{i,j}^{\rm eff} n_i(x) n_j(x)$.

In the following, in order to detect charge-density order, we compute the 
density-density structure factor (that is a diagonal operator in the 
$|x\rangle$ basis):
\begin{equation}\label{eq:Nq}
N(q) = \frac{1}{L} \sum_{i,j} \langle n_i n_j\rangle e^{i q (r_i-r_j)}.
\end{equation}
When order is present with a given periodicity $Q$, then $N(Q)/L$ is 
finite in the thermodynamic limit. Similarly, orbital order can be detected
by considering, for example, the density-density correlations of the same 
orbital on different sites:
\begin{equation}\label{eq:Nqorb}
N^\alpha(q) = \frac{1}{L} \sum_{i,j} 
\langle n^\alpha_i n^\alpha_j\rangle e^{i q (r_i-r_j)}.
\end{equation}

\begin{figure}
\includegraphics[width=\columnwidth]{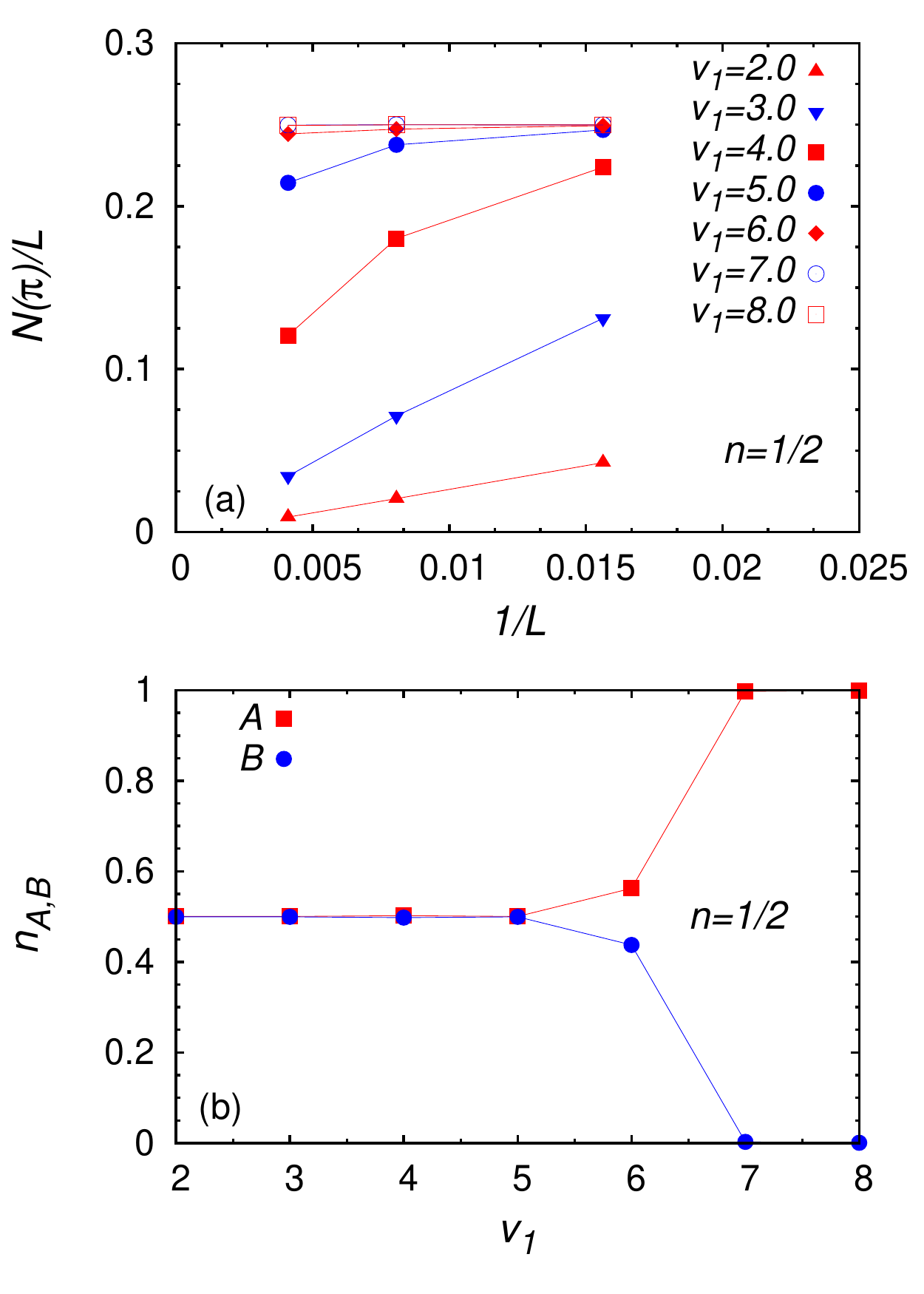}
\caption{\label{fig:1dCDW}
(Color online) (a) Size scaling of $N(Q)/L$ at $Q=\pi$ for the one-dimensional 
case at quarter filling $n=1/2$. The variational wave function is given by 
Eqs.~(\ref{eq:WF1dSlater}) and~(\ref{eq:WF1dJastrow}) with $g=10$ and 
different values of $v_1$. (b) Densities of the two sublattices 
$A$ and $B$ for the one-dimensional system at quarter filling $n=1/2$ as a 
function of the Jastrow parameter $v_1$ (with $g=10$) for $L=244$ sites.}
\end{figure}

\begin{figure}
\includegraphics[width=\columnwidth]{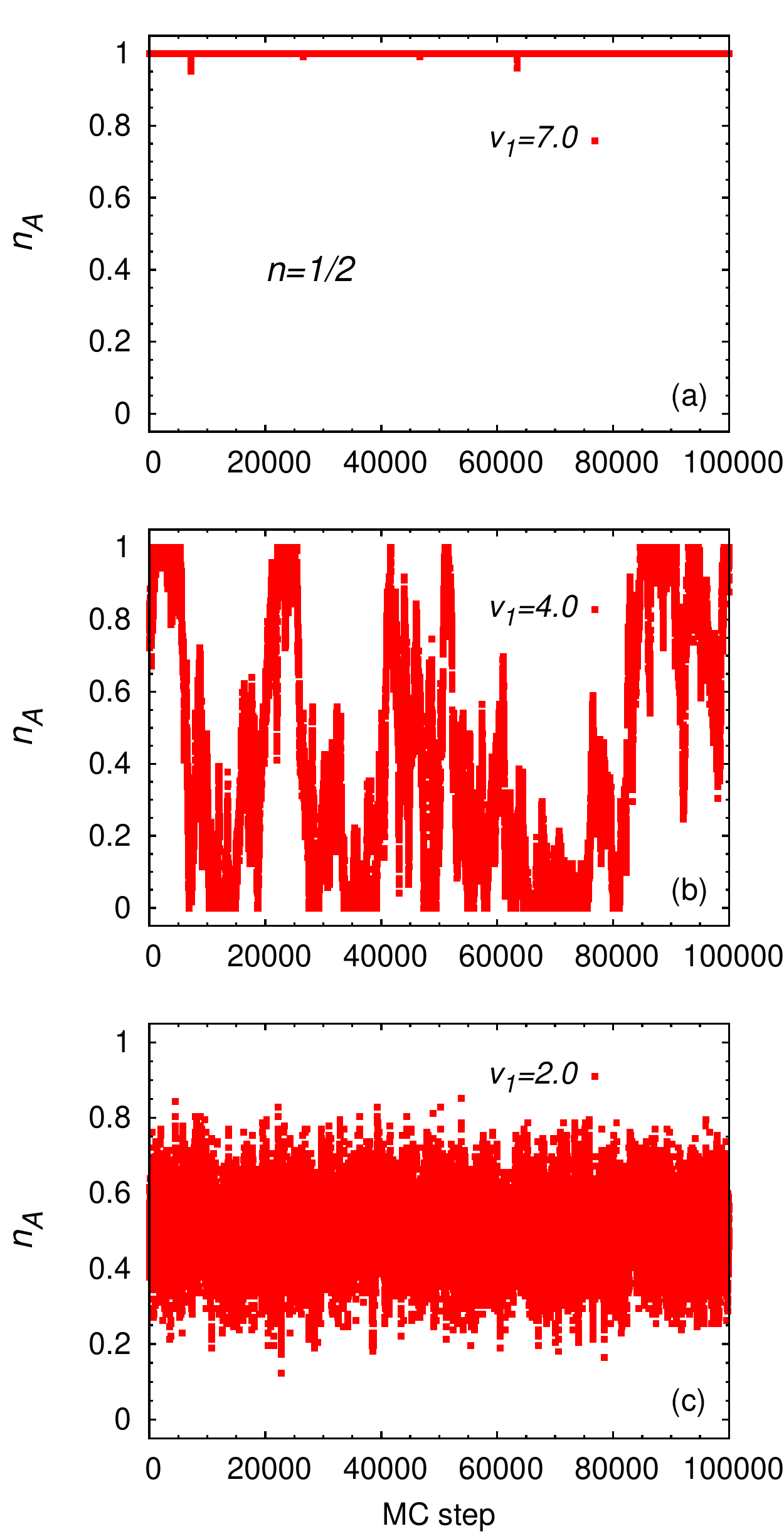}
\caption{\label{fig:1dergo}
(Color online) Monte Carlo evolution of the charge density on one sublattice 
for the one-dimensional system and $n=1/2$ for $L=244$ sites. Three different 
values of the Jastrow pseudo-potential $v_1$ are shown for the wave function 
described by Eqs.~(\ref{eq:WF1dSlater}) and~(\ref{eq:WF1dJastrow}).
$v_1=7.0$ (a), $v_1=4.0$ (b), $v_1=2.0$ (c).}
\end{figure}

\subsection{Charge-density order in one dimension}

Let us start by considering a one-dimensional system at quarter filling, i.e.,
$n=1/2$. We analyze the properties of the Jastrow-Slater wave function in
which the noninteracting state is given by filling the lowest-energy levels
of free fermions having $\epsilon(k)=-2\cos k$:
\begin{equation}\label{eq:WF1dSlater}
|\Phi\rangle = \prod_{k<k_F,\sigma} c^\dag_{k,\sigma} |0\rangle,
\end{equation}
where $k_F=\pi/4$ for quarter filling. In order to have a unique state, we 
consider chains with $L=8l+4$ sites, with $l$ integer, and periodic boundary 
conditions. In addition, we take a simple Jastrow term that only contains 
on-site and nearest-neighbor pseudo-potentials:
\begin{equation}\label{eq:WF1dJastrow} 
{\cal J} = \exp \left( -g \sum_{i} n_{i,\uparrow} n_{i,\downarrow} 
-v_1 \sum_{i} n_i n_{i+1} \right),
\end{equation}
where we fix $g=10$ and vary $v_1$. Both the noninteracting state
$|\Phi\rangle$ and the Jastrow term ${\cal J}$ are clearly invariant under 
translation and inversion symmetries. Nevertheless, the correlated wave 
function $|\Psi_J\rangle$ may describe two distinct phases for $n=1/2$: 
for small values of $v_1$, the density is uniform in the lattice (the quantum 
state is metallic), while for large values of $v_1$ there is a $1{-}0{-}1{-}0$
density order (corresponding to a charge-density-wave insulator). 
We would like to mention that the variational wave function defined by 
Eqs.~(\ref{eq:WF1dSlater}) and~(\ref{eq:WF1dJastrow}) is suitable for the
extended Hubbard Hamiltonian that includes both on-site and nearest-neighbor 
interactions~\cite{mila1993,penc1994,yoshioka2000,sano2004,ejima2005,shirakawa2009}:
\begin{eqnarray}
{\cal H} = &-&t \sum_{i,\sigma} c^\dag_{i,\sigma} c_{i+1,\sigma} + \textrm{h.c.}
             +U \sum_{i} n_{i,\uparrow} n_{i,\downarrow} \nonumber \\
           &+&V \sum_{i} n_{i} n_{i+1}.
\end{eqnarray}
The existence of a phase transition when changing $v_1$ can be understood from 
the classical mapping. When $v_1$ is large, the first term of the r.h.s. of 
Eq.~(\ref{eq:classicalpot}) dominates and drives the system into an ordered 
phase (this is expected from the classical model with nearest-neighbor 
interactions at low enough temperatures); by contrast, when $v_1$ is small, 
the first term of the r.h.s. of Eq.~(\ref{eq:classicalpot}) does not give rise 
to charge-density order (i.e., the classical temperature is large). Notice that 
in this reasoning we assume that the contribution from the Slater determinant, 
i.e., the second term of the r.h.s. of Eq.~(\ref{eq:classicalpot}), is not able 
to produce any transition, as expected for the chosen noninteracting 
part of Eq.~(\ref{eq:WF1dSlater}).

At quarter filling, the density-density structure factor~(\ref{eq:Nq}) computed
over the correlated wave function shows a peak at $Q=\pi$, which behaves 
differently for small and large values of the parameter $v_1$. 
In Fig.~\ref{fig:1dCDW}(a), we report the size scaling of $N(Q)/L$ at $Q=\pi$ 
for different values of $v_1$. Here, a drastic change can be seen when varying 
$v_1$: for $v_1 \lesssim 5$ there is no charge-density order, i.e., $N(Q)/L$ goes 
to zero in the thermodynamic limit, while for $v_1 \gtrsim 5$ there is a clear 
evidence of order, $N(Q)/L$ being finite. For $v_1 \simeq 5$ considerable size 
effects are present, as expected close to a phase transition. The averaged 
values of the densities in the two sublattices $A$ and $B$, 
$n_{A(B)}= 2/L \sum_{i \in A(B)} n_i$, as a function of $v_1$, are reported in 
Fig.~\ref{fig:1dCDW}(b).

It is important to notice that a breaking of the ergodicity (when using 
single-electron updates in the Monte Carlo sampling) is manifest when 
$v_1 \gtrsim 5$: while for small values of $v_1$ ergodicity is clearly obtained,
for large $v_1$ ergodicity is broken and the simulation remains trapped into 
one of the possible degenerate (global) minima, with specific charge patterns. 
In Fig.~\ref{fig:1dergo}, we report the evolution of the averaged charge density 
on one sublattice, as a function of Monte Carlo updating, for three values of 
$v_1$. While for $v_1=2$ [see Fig.~\ref{fig:1dergo}(c)] the charge density is 
perfectly uniform, with relatively small fluctuations around $n=1/2$, for $v_1=4$ 
[see Fig.~\ref{fig:1dergo}(b)] the evolution starts to have large oscillations 
between $0$ and $1$ (here, the two degenerate minima are already developed, but 
the barriers between them can be easily overcome); eventually, for $v_1=7$ [see 
Fig.~\ref{fig:1dergo}(a)], ergodicity is broken and the charge density remains 
stuck in one minimum, since single-electron moves do not allow the system to 
tunnel easily to the other minimum. We mention that the large Gutzwiller factor 
used in the calculation prevents the density to be larger than 1 on each site, 
as it is clear from Fig.~\ref{fig:1dergo}.

\begin{figure}
\includegraphics[width=\columnwidth]{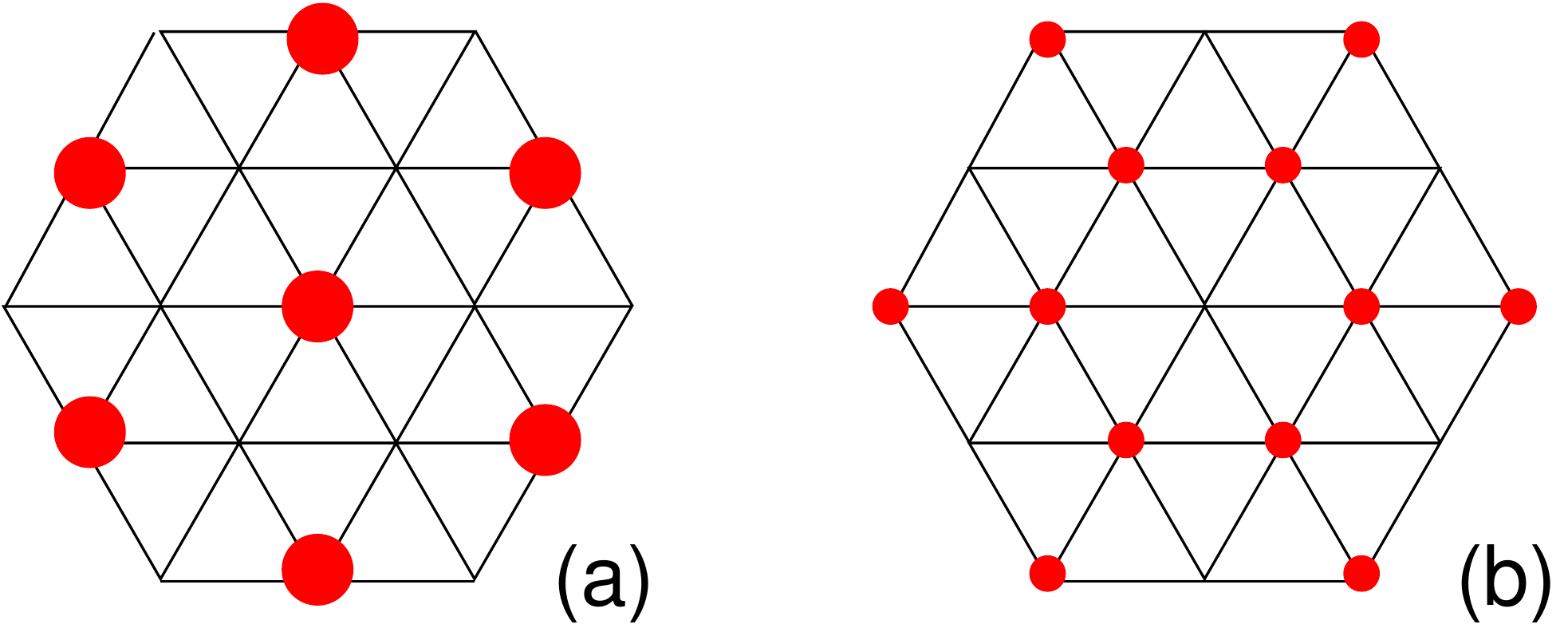}
\caption{\label{fig:2dpicture}
(Color online) Cartoon picture of the $2{-}0{-}0$ (a) and $1{-}1{-}0$ (b)
phases that can be stabilized in the triangular lattice at $n=2/3$ 
filling, when considering the wave function of Eqs.~(\ref{eq:WF2dSlater})
and~(\ref{eq:WF2dJastrow}).}
\end{figure}

\subsection{Charge-density order in the triangular lattice}

Charge-density order can be easily obtained also in two spatial dimensions. 
As an example, we consider the case of a triangular lattice with 
$n=2/3$~\cite{watanabe2005}. We take, similarly to the one-dimensional case, 
a Slater part in which the lowest-energy levels of a free-fermion Hamiltonian 
are filled, e.g., $\epsilon(k)=-2[\cos k_x + \cos (k_x/2 + \sqrt{3} k_y/2) +
\cos(k_x/2 - \sqrt{3} k_y/2)]$:
\begin{equation}\label{eq:WF2dSlater}
|\Phi\rangle = \prod_{k<k_F,\sigma} c^\dag_{k,\sigma} |0\rangle.
\end{equation}
Then, we consider a Jastrow term that contains on-site and nearest-neighbor 
terms:
\begin{equation}\label{eq:WF2dJastrow} 
{\cal J} = \exp \left( -g \sum_{i} n_{i,\uparrow} n_{i,\downarrow} 
         -v_1 \sum_{\langle i,j \rangle} n_i n_{j} \right),
\end{equation}
where $\langle \dots \rangle$ indicates the nearest-neighbor bonds of the lattice; 
both $g$ and $v_1$ are variational parameters that are varied. As for the 
one-dimensional case discussed before, the variational wave function defined by
Eqs.~(\ref{eq:WF2dSlater}) and~(\ref{eq:WF2dJastrow}) is suitable for the extended
Hubbard model with both on-site $U$ and nearest-neighbor $V$ interactions.
In this case, we can describe three different phases: the first one, with small 
$g$ and $v_1$, has uniform densities (corresponding to a metal), the second one, 
with small $g$ and large $v_1$, develops a charge-density order in which a site 
with $2$ electrons is surrounded by empty sites [denoted by $2{-}0{-}0$ order;
this notation indicates the number of electrons in a triangle; see 
Fig.~\ref{fig:2dpicture}(a)], and the third one, with large $g$ and small $v_1$, 
has another kind of charge-density order in which one empty site is surrounded by 
singly-occupied sites [denoted by $1{-}1{-}0$ order; see 
Fig.~\ref{fig:2dpicture}(b)]. As before, this scenario can be understood from the 
classical mapping of Eq.~(\ref{eq:classicalpot})~\cite{notemotrunich}.

\begin{figure}
\includegraphics[width=\columnwidth]{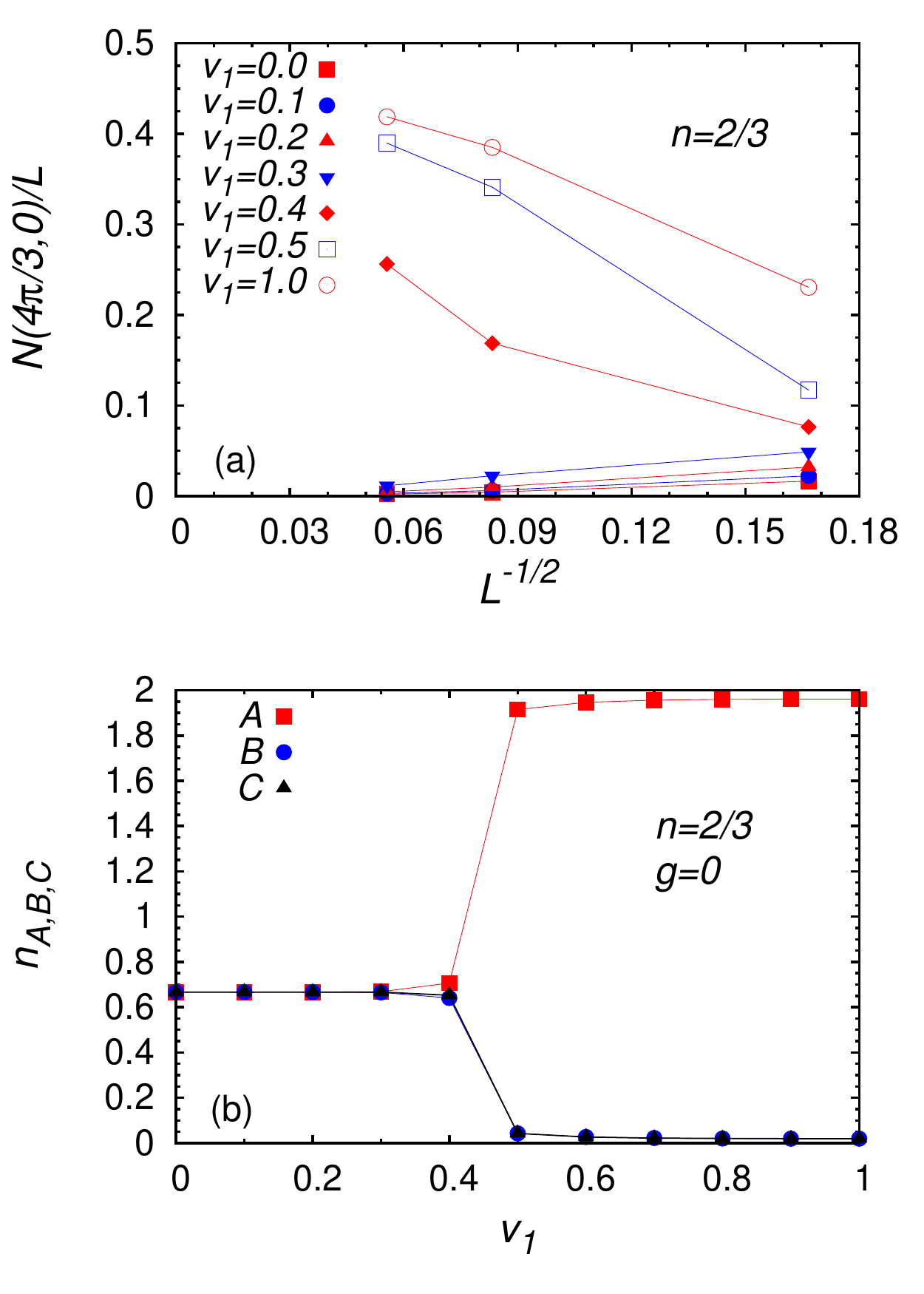}
\caption{\label{fig:2dCDW_Nq}
(Color online) (a) Size scaling of the density-density structure factor 
$N(Q)/L$ of Eq.~(\ref{eq:Nq}) at $Q =(4\pi/3,0)$ for the triangular lattice and 
filling $n=2/3$. The variational wave function is given by 
Eqs.~(\ref{eq:WF2dSlater}) and~(\ref{eq:WF2dJastrow}), with $g=0$ and different 
values of $v_1$. (b) Densities of the three sublattices $A$, $B$, and $C$, as a 
function of the Jastrow parameter $v_1$, for the same wave function of the upper 
panel, on a cluster with $L=324$ sites.}
\end{figure}

Let us start by considering $g=0$ and varying the nearest-neighbor parameter
$v_1$. In this way, we can have a transition between a phase with uniform 
densities and another phase with $2{-}0{-}0$ order. In fact, the size scaling 
of the density-density correlation function~(\ref{eq:Nq}) for $Q=(4\pi/3,0)$ 
[or the symmetry-related one $Q=(2\pi/3,2\pi/\sqrt{3})$] shows clear evidence
of order in the thermodynamic limit for $v_1 \gtrsim 0.4$; see 
Fig.~\ref{fig:2dCDW_Nq}(a). Correspondingly, the local densities on the three 
sublattices acquire different values when $v_1 \gtrsim 0.4$; see 
Fig.~\ref{fig:2dCDW_Nq}(b).

\begin{figure}
\includegraphics[width=\columnwidth]{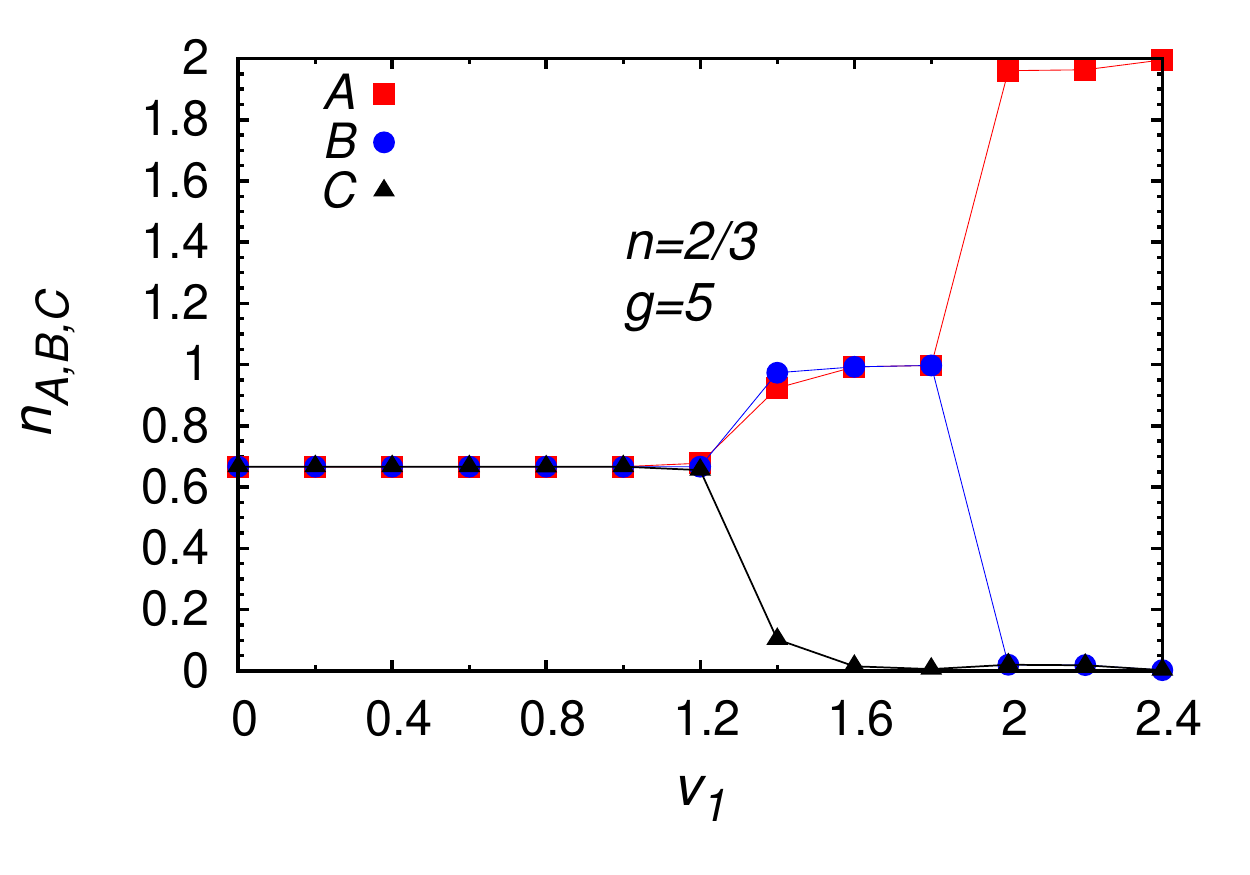}
\caption{\label{fig:2dCDW_density}
(Color online) Densities of the three sublattices $A$, $B$, and $C$ for the 
triangular lattice at $n=2/3$ as a function of the Jastrow parameter $v_1$, for 
the variational wave function given by Eqs.~(\ref{eq:WF2dSlater}) 
and~(\ref{eq:WF2dJastrow}) with $g=5$; the cluster has $L=324$ sites.}
\end{figure}

A richer scenario appears when the Gutzwiller factor $g$ is finite. Indeed,
the effect of $g$ is to suppress doubly-occupied sites and, therefore, it
acts against the $2{-}0{-}0$ phase, favoring instead the $1{-}1{-}0$ order.
In Fig.~\ref{fig:2dCDW_density}, we report the densities on the three 
sublattices for the case where $g=5$ and $v_1$ is varied from $0$ to $2.4$. 
The effect of the on-site Jastrow term is clear: on the one hand, it enlarges 
the stability of the uniform phase, up to $v_1 \simeq 1.2$; on the other hand, 
it creates an intermediate phase in which two sites have $n_i \approx 1$ and 
another one has $n_i \approx 0$ (the spatial pattern is such that the empty 
site is surrounded by occupied sites). Then, for a large enough nearest-neighbor
Jastrow parameter, i.e., $v_1 \gtrsim 2$, the $2{-}0{-}0$ state is obtained 
again.

\subsection{Orbital order in a two-band model}

Let us now turn to a two-band model and show that a simple choice of the
Jastrow factor~(\ref{eq:jastrowmo}) may give rise to orbital order. We focus 
our attention on the two-dimensional square lattice at half-filling, $n=2$ 
(i.e., two electrons per site, each site having two orbitals). The Slater part 
is constructed from two bands having different width, e.g., 
$\epsilon_1(k)=-2(\cos k_x +\cos k_y)$ and 
$\epsilon_2(k)=-(\cos k_x +\cos k_y)$, by filling the lowest-energy states:
\begin{equation}\label{eq:WForbSlater}
|\Phi\rangle = \prod_{k<k_F,\alpha,\sigma} c^\dag_{k,\alpha,\sigma} 
|0\rangle,
\end{equation}
where $\alpha=1$, $2$ indicates the two bands.

The Jastrow factor contains both on-site intra- and inter-orbital terms and
the nearest-neighbor intra-orbital term:
\begin{equation}\label{eq:WForbJastrow}
{\cal J} = \exp \left( 
- \frac{1}{2} \sum_{i,\alpha,\beta} g^{\alpha,\beta} n^\alpha_i n^\beta_i
- \sum_{\langle i,j\rangle,\alpha} v^{\alpha,\alpha}_1 
n^\alpha_i n^\alpha_j \right),
\end{equation}
where $\langle \dots \rangle$ indicates the nearest-neighbor bonds of the lattice. 
In the following, we will fix $g_{1,1}=g_{2,2}=2$ and $g_{1,2}=g_{2,1}=1$ and vary 
$v^{1,1}_1=v^{2,2}_1$. The wave function defined by Eqs.~(\ref{eq:WForbSlater}) 
and~(\ref{eq:WForbJastrow}) is suitable to describe the phases of a two-band 
Hubbard model with both intra-band ($U$) and inter-band ($U^\prime$) interactions
for $U<U^\prime$ and even for $U=U^\prime$ within the paramagnetic 
sector~\cite{tocchio2015}:
\begin{eqnarray}
{\cal H} = &-& \sum_{\alpha} t_{\alpha} \sum_{\langle i,j \rangle,\sigma}
             c^\dag_{i,\alpha,\sigma} c_{j,\alpha,\sigma} + \textrm{h.c.} \nonumber \\
           &+& U \sum_{i,\alpha} n_{i,\alpha,\uparrow} n_{i,\alpha,\downarrow} +
               U^\prime \sum_{i} n_{i,1} n_{i,2}.
\end{eqnarray}

\begin{figure}
\includegraphics[width=0.8\columnwidth]{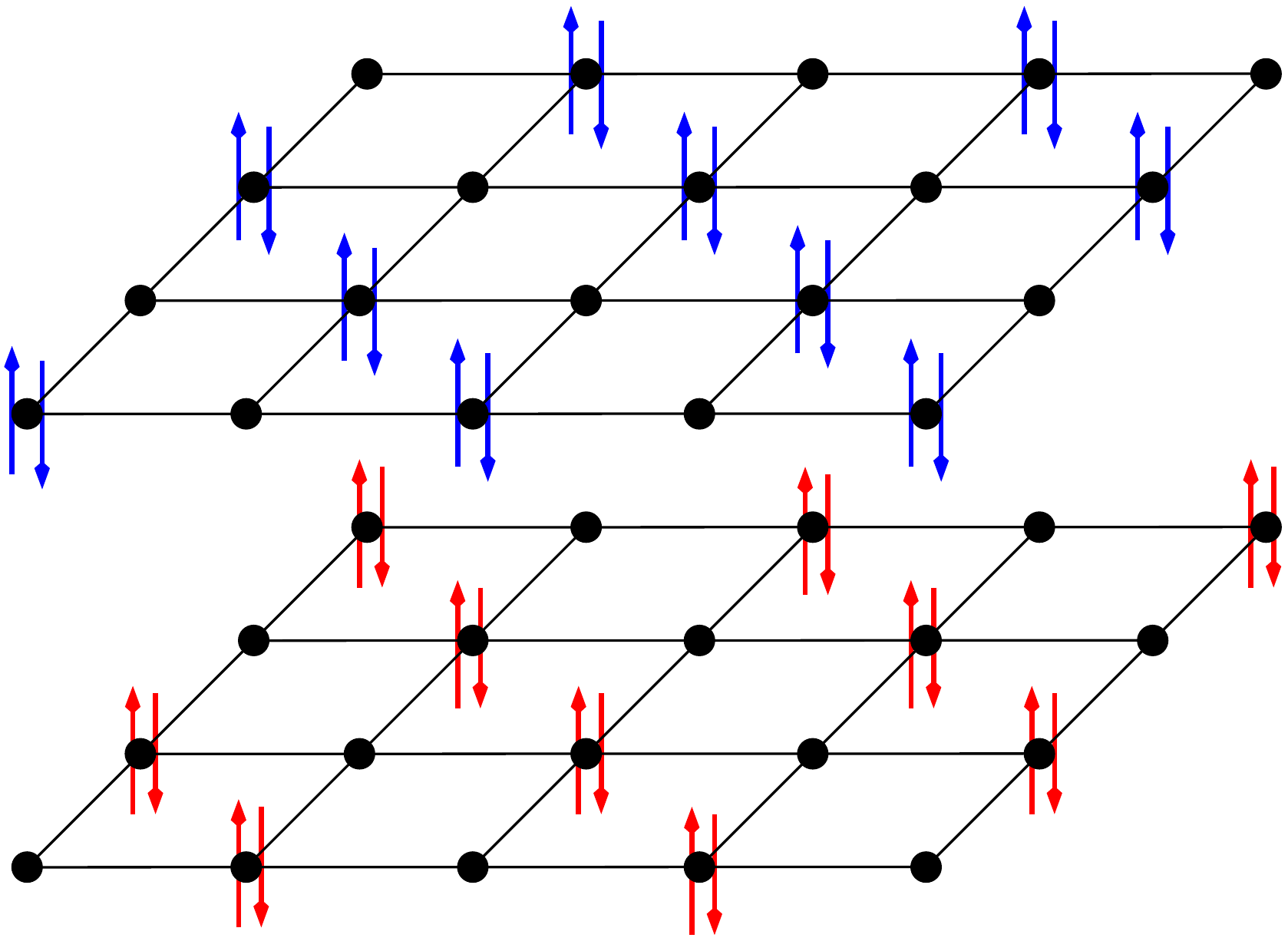}
\caption{\label{fig:pictureorb}
(Color online) Cartoon picture of the orbital-ordered phase that can be 
stabilized in the square lattice at $n=2$ filling, when considering the wave 
function of Eqs.~(\ref{eq:WForbSlater}) and~(\ref{eq:WForbJastrow}). The two
orbitals are shown as different layers.}
\end{figure}

\begin{figure}
\includegraphics[width=\columnwidth]{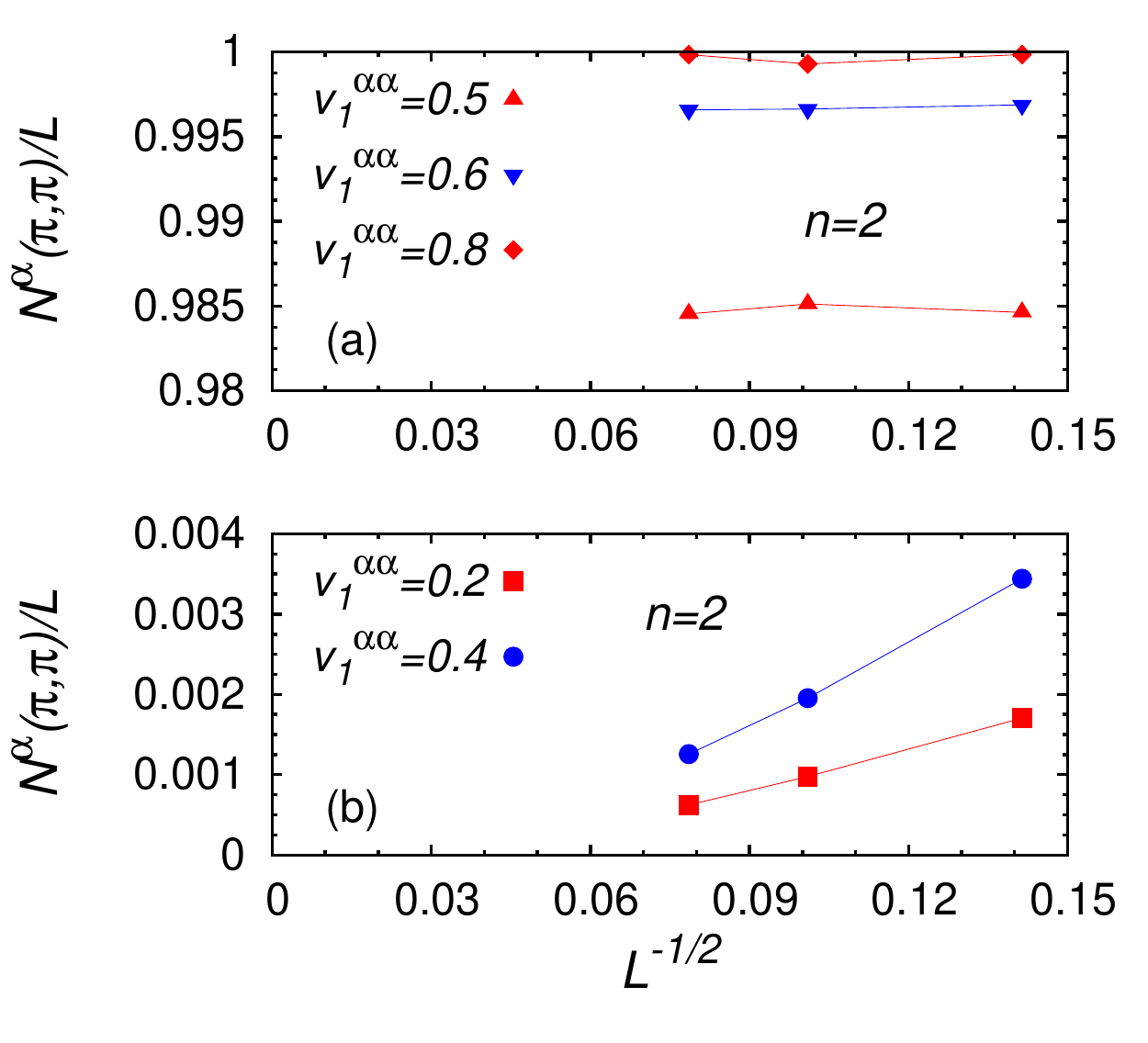}
\caption{\label{fig:orbital}
(Color online) Size scaling of the density-density structure factor 
$N^\alpha(Q)/L$ of Eq.~(\ref{eq:Nqorb}) with $\alpha=1$ for the two-band model
on the square lattice at filling $n=2$. The variational wave function is given 
by Eqs.~(\ref{eq:WForbSlater}) and~(\ref{eq:WForbJastrow}), with 
$g_{1,1}=g_{2,2}=2$ and $g_{1,2}=g_{2,1}=1$. The values of 
$v_1^{\alpha,\alpha}=0.5$, $0.6$, and $0.8$ are shown in (a), while 
$v_1^{\alpha,\alpha}=0.2$ and $0.4$ are shown in (b).}
\end{figure}

As a function of $v_1^{\alpha,\alpha}$, the wave function describes a transition
from a state with uniform densities on each orbital to a state with orbital order,
in which the two electrons per site reside on the same orbital (with opposite 
spin), two neighboring sites having different orbitals occupied; see 
Fig.~\ref{fig:pictureorb}. This symmetry-broken state can be achieved for a
sufficiently large value of $v_1^{\alpha,\alpha}$. In Fig.~\ref{fig:orbital}, 
we show the size scaling of the orbital-resolved density-density structure 
factor of Eq.~(\ref{eq:Nqorb}) for $\alpha=1$ and $Q=(\pi,\pi)$. 
For $v_1^{\alpha,\alpha} \lesssim 0.4$, the size scaling clearly indicates 
that the wave functions has no orbital order, while for 
$v_1^{\alpha,\alpha} \gtrsim 0.5$, $N^\alpha(Q)/L$ is finite in the 
thermodynamic limit, implying orbital order.

We finally mention that, as shown in Ref.~\onlinecite{tocchio2015}, orbital 
order can be also favored by the presence of an on-site intra-band pairing; 
however, here we preferred to consider the simple Slater determinant of 
Eq.~(\ref{eq:WForbSlater}) and demonstrate that orbital order can be achieved 
by the Jastrow term~(\ref{eq:WForbJastrow}) only.

\begin{figure*}
\includegraphics[width=0.9\textwidth]{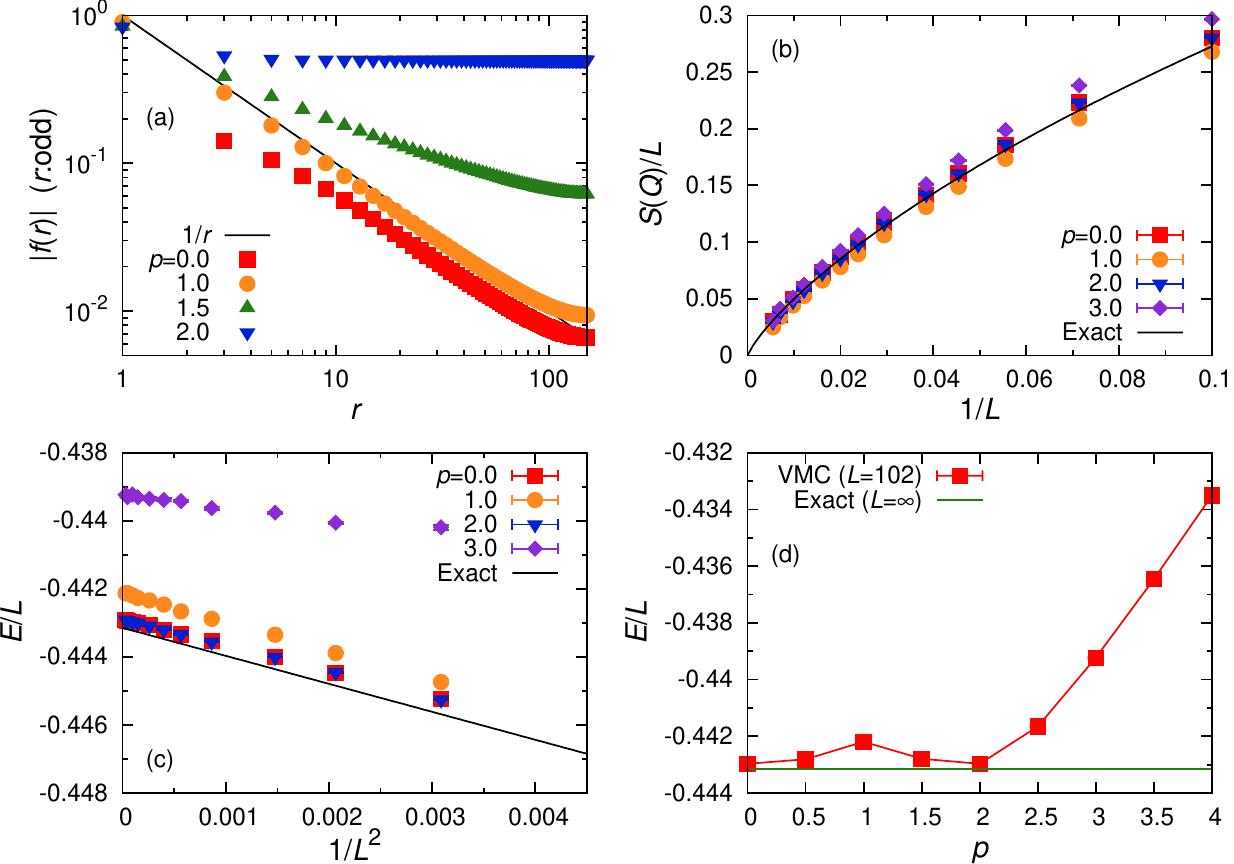}
\caption{\label{fig:1dcase}
(Color online) Results for the wave function described by 
Eqs.~(\ref{eq:gapless1da}) and~(\ref{eq:gapless1db}) for the one-dimensional
lattice. (a) Real-space behavior of the absolute value of the pairing 
amplitude $f(r)$ as a function of the distance $r$ (for opposite sublattices).
(b) Size scaling of the spin-spin structure factor $S(Q)/L$ for $Q=\pi$; 
the exact (at the leading-order) size scaling for the Heisenberg model
is also reported for comparison~\cite{hallberg1995}. (c) Size scaling of the 
energy per site $E/L$; the exact (up to order $1/L^2$) size scaling for the 
Heisenberg model is also reported for comparison~\cite{affleck1986}.
(d) Energy per site $E/L$ versus the parameter $p$ of Eq.~(\ref{eq:gapless1db}) 
for $L=102$; the exact value of the Heisenberg model for the thermodynamic 
limit is also reported for comparison.}
\end{figure*}

\section{Magnetic and dimer ordering}\label{sec:magnetic_order}

\subsection{General concepts and magnetic order}

Let us first focus on the possibility of having magnetic long-range 
order in the fully projected wave function~(\ref{eq:gutzwiller}). 
Of course, magnetic order is certainly present whenever the 
noninteracting state $|\Phi\rangle$ is obtained from an 
uncorrelated Hamiltonian that {\it explicitly} contains a
magnetic order parameter, thus breaking the spin SU(2)
symmetry~\cite{yokoyama1987c,chen1990,becca2011}. This is a trivial 
case that will not be considered here. Instead, we focus on the more 
interesting case in which $|\Phi\rangle$ has no magnetic order.

Indeed, for certain choices of $|\Phi\rangle$, the Gutzwiller 
projector of Eq.~(\ref{eq:fullgutz}) may generate long-range order. 
The fully projected wave function may be written in term of a linear 
superposition of singlet coverings of the lattice~\cite{anderson1987}, 
the difference with respect to the bosonic LDA state residing upon the 
actual values of the amplitudes of various singlet coverings. Since, 
in general, there is not a one-to-one relation between bosonic
and fermionic representations of the RVB states~\cite{becca2011}, it is not
{\it a-priori} obvious that fermionic states may describe magnetically 
ordered states, as the LDA wave function does. 

To have a transparent RVB representation, we use the following
parametrization of the noninteracting 
state~\cite{himeda2000,tahara2008}:
\begin{equation}\label{eq:psifij}
|\Phi\rangle = \exp \left( 
               \sum_{i,j} f_{i,j} c^\dag_{i,\uparrow} c^\dag_{j,\downarrow}
               \right) |0\rangle,
\end{equation}
which can be obtained as the ground state of a BCS Hamiltonian, containing
both pairing and hopping (without performing particle-hole transformations).
Here, $c^\dag_{i,\sigma}$ creates an electron on site $i$ with spin $\sigma$.
Then, since Eq.~(\ref{eq:psifij}) does not conserve the number of particles,
the correlated state $|\Psi\rangle$ of Eq.~(\ref{eq:gutzwiller}) 
must involve a further projection ${\cal P}_N$ on the subspace with $N=L$ 
particles.

In Eq.~(\ref{eq:psifij}), $f_{i,j}$ is the pair amplitude, which is taken to 
be symmetric to form singlets in the $(i,j)$ bond:
\begin{equation}
f_{i,j} = f_{j,i};
\end{equation}
in this way, $|\Phi\rangle$ is a total singlet and does not break spin SU(2)
symmetry. Moreover, we consider pairing functions that have all the lattice 
symmetries. Therefore, as for the LDA wave function, the pairing amplitude 
$f(r)$ only depends upon the bond length $r$ (bonds with the same $r$ may 
have different $f(r)$ whenever they are not related by point-group symmetries).
The Fourier transform of $f(r)$ is denoted by $f(k)$.

In contrast to charge or orbital order, which are mainly driven by the Jastrow 
factor~(\ref{eq:jastrow}), the appearance of magnetic order cannot be easily 
explained through a classical mapping, e.g., Eqs.~(\ref{eq:classical})
and~(\ref{eq:classicalpot}). Instead, it is mainly due to two circumstances:
(i) the presence of the full Gutzwiller projector that enforces no double
occupation and (ii) the presence of long-range singlets that create a strong 
entanglement among spins at very large distances.

Here, we consider one-dimensional chains and the two-dimensional 
square lattice. Similarly to what has been demonstrated within 
the bosonic representation of the RVB state, we expect that 
magnetic order may appear whenever the fully projected state has 
the Marshall signs and the pairing amplitude decays sufficiently slowly, i.e., 
$f(r) \propto 1/r^\alpha$ with a small $\alpha$. 

Explicitly, we consider a specific parametrization of the pairing
amplitude:
\begin{equation}\label{eq:fkbcs}
f(k) = \frac{\Delta(k)}{\epsilon(k)+\sqrt{\epsilon^2(k)+\Delta^2(k)}},
\end{equation}
which results from considering $|\Phi\rangle$ as the ground state of the BCS 
Hamiltonian:
\begin{equation}\label{eq:bcs}
{\cal H}_{BCS} = \sum_{k,\sigma} \epsilon(k) c^\dag_{k,\sigma} c^\dagga_{k,\sigma} 
               + \sum_{k} \Delta(k) c^\dag_{k,\uparrow} c^\dag_{-k,\downarrow}
               + \textrm{h.c.},
\end{equation}
where $c^\dag_{k,\sigma}$ ($c^\dagga_{k,\sigma}$) creates (destroys) an 
electron with momentum $k$ and spin $\sigma$ (along the $z$ axis); 
$\Delta(k)=\Delta(-k)$ is the singlet pairing amplitude. The BCS spectrum is 
given by:
\begin{equation}
E(k) = \pm \sqrt{\epsilon^2(k)+\Delta^2(k)}.
\end{equation}
A gapless (gapped) BCS spectrum $E(k)$ corresponds to a power-law (exponential) 
decay of the pairing function $f(r)$. Since for the bosonic LDA wave function 
the existence of magnetic order is related to a sufficiently slow decay of the 
pairing function, we expect that a gapped BCS spectrum does not give rise to 
magnetic order.

To fulfill the Marshall sign rule, it is sufficient to 
take~\cite{becca2011}:
\begin{eqnarray}
\epsilon(k+Q) &=& -\epsilon(k), \label{eq:marshall1} \\
\Delta(k+Q) &=& -\Delta(k), \label{eq:marshall2}
\end{eqnarray}
with $Q=\pi$ in one dimension and $Q=(\pi,\pi)$ for the square lattice.
Given the definition of the pairing function~(\ref{eq:fkbcs}), we have:
\begin{equation}\label{eq:marshall}
f(k+Q) = -\frac{1}{f(k)}.
\end{equation}

In the following, we will investigate the possibility of having long-range 
magnetic order with the constraint of Eq.~(\ref{eq:marshall}) by varying 
the exponent $\alpha$ of the power-law decay $f(r) \propto 1/r^\alpha$.
Magnetic order can be detected by evaluating the spin-spin structure factor:
\begin{equation}\label{eq:Sq}
S(q) = \frac{1}{L} \sum_{i,j} \langle \boldsymbol{S}_i\cdot\boldsymbol{S}_j \rangle
e^{i q \cdot (r_i-r_j)};
\end{equation}
magnetic order with a given pitch vector $Q$ is present whenever the moment 
(squared) $m^2=S(Q)/L$ is finite in the thermodynamic limit.

\begin{figure}
\includegraphics[width=\columnwidth]{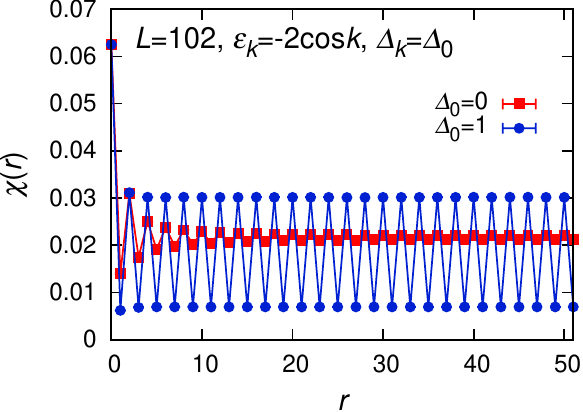}
\caption{\label{fig:dimerization}
(Color online) Dimer-dimer correlations of Eq.~(\ref{eq:dimerdimer}) for
the one-dimensional wave function obtained with Eqs.~(\ref{eq:gapped1da})
and~(\ref{eq:gapped1db}). The gapless case has $\Delta_0=0$, while the gapped 
one has $\Delta_0=1$.}
\end{figure}

\begin{figure}
\includegraphics[width=\columnwidth]{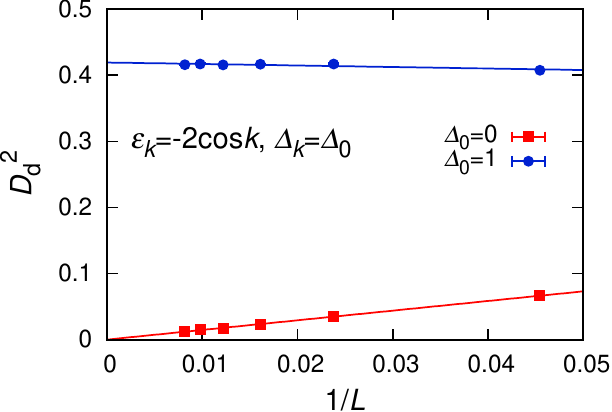}
\caption{\label{fig:sizescalingdim}
(Color online) Size scaling of the square of the dimer order parameter of
Eq.~(\ref{eq:dimerord}) for the gapless ($\Delta_0=0$) and gapped 
($\Delta_0=1$) cases.}
\end{figure}

\subsection{RVB wave functions in one dimension}

In one spatial dimension (and short-range interactions), antiferromagnetic
order is forbidden by the Mermin-Wagner theorem both in the ground state 
and at finite temperature~\cite{pitaevskii1991}. Nevertheless, variational 
wave functions may possess long-range order, as a matter of principle. 
This is, e.g., the case for bosonic RVB states, as shown in 
Ref.~\onlinecite{lin2012}. 

In the following, we consider the parametrization~(\ref{eq:fkbcs}) with:
\begin{eqnarray}
\label{eq:gapless1da}
\epsilon(k) &=& -2\cos k, \\
\label{eq:gapless1db}
\Delta(k) &=&
\left\{
\begin{array}{cc}
 +|\epsilon(k)|^p & {\rm for} \;\; \epsilon(k)<0, \\
 -|\epsilon(k)|^p & {\rm for} \;\; \epsilon(k)>0,
\end{array}
\right.
\end{eqnarray}
that allows us to easily control the exponent of the power-law decay of $f(r)$. 
This Ansatz obeys the Marshall sign rule, as clearly seen from 
Eqs.~(\ref{eq:marshall1}) and~(\ref{eq:marshall2}). We emphasize that this 
parametrization contains the case of a free Fermi sea (where all states below 
$k_F$ are occupied while the others are empty) that can be obtained by taking 
$p=1$:
\begin{equation}\label{eq:fermisea}
 f(k) =
\left\{
\begin{array}{cc}
 1 & {\rm for} \;\; |k|<k_F, \\
 0 & {\rm for} \;\; |k|>k_F.
\end{array}
\right.
\end{equation}
Within the parametrization given by Eqs.~(\ref{eq:gapless1da}) 
and~(\ref{eq:gapless1db}), we have that the long-range behavior for $f(r)$ is 
given by $f(r) \propto 1/r^\alpha$, with $\alpha=1$ for $p \le 1$, $\alpha=2-p$
for $1 \le p \le 2$, and $\alpha=0$ for $p \ge 2$ (the latter case implying 
that $f(r)$ approaches a constant for large $r$). In Fig.~\ref{fig:1dcase}(a), 
we report the results of the pairing amplitude for different values of $p$. 
The spin-spin structure factor shows a peak at $Q=\pi$; however, in all cases, 
the wave function does not possess magnetic long-range order, in agreement with
the Mermin-Wagner theorem, since $m$ vanishes in the thermodynamic limit, as 
shown in Fig.~\ref{fig:1dcase}(b). Moreover, also the leading-order corrections
in the system size are correct for any value of $\alpha$, i.e., 
$S(Q) \propto (\ln c L/2)^{3/2}$, where $c=25.5$~\cite{hallberg1995}.

In the following, we would like to discuss the accuracy of this class of wave 
functions for the unfrustrated Heisenberg model:
\begin{equation}
{\cal H} = J \sum_{i} \boldsymbol{S}_i\cdot\boldsymbol{S}_{i+1}.
\end{equation}
It is well known~\cite{gros1987} that already the fully projected Fermi 
sea~(\ref{eq:fermisea}) represents a very good variational ansatz for this 
model, with an accuracy on the ground-state energy of about $0.2\%$ [i.e., 
$E/J=-0.44212(1)$ compared with the exact value $E_{ex}/J=1/4-\ln2=-0.44315$].
Within this class of states, we can strongly improve the accuracy of the 
fully projected Fermi sea: the best energies are obtained for $p=0$ 
[$E/J=-0.44290(1)$] and $p=2$ [$E/J=-0.44289(1)$]; see Fig.~\ref{fig:1dcase}(d). 
In all cases, the finite-size scaling of the energy per site shows the correct 
behavior in the leading-order corrections (up to order $1/L^2$)~\cite{affleck1986}; 
see Fig.~\ref{fig:1dcase}(c). 

\begin{figure*}
\includegraphics[width=0.9\textwidth]{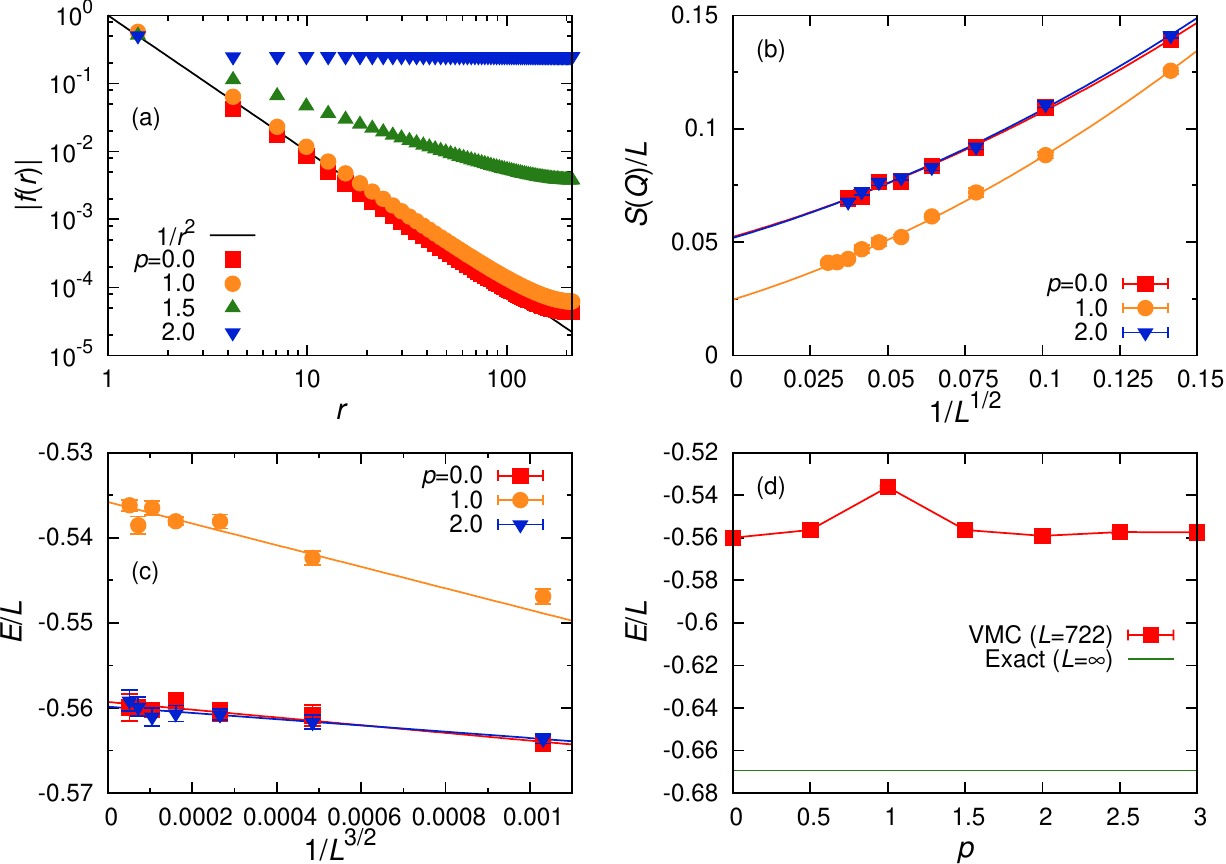}
\caption{\label{fig:2dcase}
(Color online) The same as in Fig.~\ref{fig:1dcase} but for the two-dimensional
case. The variational wave function is described by Eqs.~(\ref{eq:para2da})
and~(\ref{eq:para2db}). In (a) $|f(r)|$ is shown along the diagonal direction
of the square lattice.}
\end{figure*}

\subsection{Dimerization}

To conclude the one-dimensional case, we consider the case of dimerization.
The possible emergence of valence-bond solids has been discussed in depth
for one- and two-dimensional spin models~\cite{read1989}. Few works
have used fermionic wave functions that explicitly break the translational
symmetry to assess the possible emergence of dimer order in various 
lattices~\cite{heidarian2009,iqbal2011,kaneko2015}. However, here we are
interested in the case where the variational wave function preserves all
symmetries, similarly to what has been done in bosonic RVB sates~\cite{lin2012}.
Indeed, dimer order is present within fermionic RVB wave functions that have 
short-range pairing amplitudes, as is typical for a gapped BCS spectrum:
\begin{eqnarray}
\label{eq:gapped1da}
\epsilon(k) &=& -2\cos k, \\
\label{eq:gapped1db}
\Delta(k) &=& \Delta_0.
\end{eqnarray}
A gapped BCS spectrum can be obtained for $\Delta_0>0$. Similarly, one could
consider $\Delta(k)=\Delta_2 \cos (2k)$ (not shown here). Both $\Delta_0$ 
and $\Delta_2$ are variational parameters. Notice that, in both cases, 
the Marshall sign rule does not apply, as expected from a generic dimerized 
phase.

For the dimer-dimer correlation function, one can consider the simplified
form that includes only $z{-}z$ correlations:
\begin{equation}\label{eq:dimerdimer}
\chi(r)= \frac{1}{L} \sum_i 
         \langle S^z_{i} S^z_{i+1} S^z_{i+r} S^z_{i+1+r} \rangle.
\end{equation}
The order parameter for long-ranged dimerization can then be defined as:
\begin{equation}\label{eq:dimerord}
D^2_d = 9 \lim_{r \to \infty} |2 \; \chi(r) - \chi(r+1) - \chi(r-1)|,
\end{equation}
where the factor $9$ is introduced in order to take into account the three
spin components.

In Fig.~\ref{fig:dimerization}, we show the dimer-dimer correlations $\chi(r)$ 
for two cases with $\Delta_0=0$ (gapless) and $\Delta_0=1$ (gapped). For the 
former case, $\chi(r) \to {\rm const}$ for large distances (in the definition 
of the dimer-dimer correlation, we do not subtract the disconnected terms), 
indicating that the wave function does not possess any dimer order. 
By contrast, for the latter case, $\chi(r)$ oscillates between two different 
values, which is the expected behavior for a dimerized system. We would like 
to mention that, in presence of a gapped BCS spectrum, both periodic and 
antiperiodic conditions can be chosen in the BCS Hamiltonian~(\ref{eq:bcs}), 
still having a unique ground state. The results shown in 
Fig.~\ref{fig:dimerization} have been obtained with periodic boundary
conditions, but a similar outcome can also be obtained with antiperiodic ones. 
These two states have momentum $k=0$ and $k=\pi$ and are 
the ones that become degenerate in the thermodynamic 
limit~\cite{becca2011,sorella2003}. The size scaling of the dimer order 
parameter~(\ref{eq:dimerord}) confirms the possibility of describing a finite 
dimerization within the class of translationally invariant (gapped) states 
of Eqs.~(\ref{eq:gapped1da}) and~(\ref{eq:gapped1db}), as shown in 
Fig.~\ref{fig:sizescalingdim}. 

\subsection{RVB wave functions in the two-dimensional square lattice}

We now discuss the possible emergence of magnetic order in the two-dimensional 
square lattice. In order to reduce the finite-size effects, we consider 
$45^\circ$ degree tilted square lattices, with $L=2l^2$ sites, $l$ being an odd
integer. Similarly to the one-dimensional case, we adopt the following 
parametrization for the pairing amplitude of Eq.~(\ref{eq:fkbcs}):
\begin{eqnarray}
\label{eq:para2da}
\epsilon(k) &=& -2(\cos k_x + \cos k_y), \\
\label{eq:para2db}
\Delta(k) &=&
\left\{
\begin{array}{cc}
 +|\epsilon(k)|^p & {\rm for} \;\; \epsilon(k)<0, \\
 -|\epsilon(k)|^p & {\rm for} \;\; \epsilon(k)>0.
\end{array}
\right.
\end{eqnarray}
As before, the projected Fermi sea is recovered with $p=1$. The dominant 
pairing amplitudes are aligned, in real space, along the diagonals, scaling as 
$f(r) \propto 1/r^\alpha$, with $\alpha=2$ for $p \le 1$, $\alpha=4-2p$ for 
$1 \le p \le 2$, and $\alpha=0$ for $p \ge 2$. As for the LDA wave function,
antiferromagnetic order is expected whenever the pairing function $f(r)$ decays
slowly with the distance. Within our parametrization, $\alpha \le 2$, which
fulfills this requirement. In Fig.~\ref{fig:2dcase}(a), we report the pairing 
function along the diagonal direction for a few values of $p$. We find that $S(q)$ 
has a peak at $Q=(\pi,\pi)$. As it was pointed out in the variational Monte Carlo 
study of Ref.~\onlinecite{li2013}, the projected Fermi sea on the square lattice 
possesses long-range magnetic order. 
However, in Ref.~\onlinecite{li2013} the actual values of the spin-spin 
correlations must be corrected by a factor $3/4$, given the definition of the 
isotropic spin-spin correlations. This fact implies that the correct value 
$m \approx 0.161$ is slightly smaller than the one reported in 
Ref.~\onlinecite{li2013}. Our data are in perfect agreement with 
$m \approx 0.161$, as shown in Fig.~\ref{fig:2dcase}(b), for the $p=1$ case.

Remarkably, long-range magnetic order is obtained for all values of $p$ within 
the parametrization of Eqs.~(\ref{eq:para2da}) and~(\ref{eq:para2db}); see 
Fig.~\ref{fig:2dcase}(b). Moreover, the actual values of the finite-size 
magnetization, as well as its thermodynamic extrapolation, are similar for 
small and large values of $p$: for example, we obtain the same values (within
a few error bars) for $p=0$ and $p=2$. In these cases, the thermodynamic 
extrapolation gives $m \approx 0.224$, substantially above the value obtained 
with $p=1$, but still below the exact value of the unfrustrated Heisenberg model
for which $m\approx 0.307$~\cite{sandvik1997,calandra1998,jiang2011}. 
Nevertheless, the fully projected wave function that we have considered here 
represents a clear example in which it is possible to realize a symmetry breaking 
within a state that preserves all the symmetries.

Furthermore, in the two-dimensional case the size effects of the energy per 
site are similar to the ones of two-dimensional ordered antiferromagnets, 
i.e., with $1/L^{3/2}$ corrections~\cite{neuberger1989,fisher1989} 
[Fig.~\ref{fig:2dcase}(c)]. However, in this case, the accuracy on the energy 
is much worse compared to the one-dimensional case, being $16\%$ for the best 
case~\cite{sandvik1997,calandra1998}; see Fig.~\ref{fig:2dcase}(d).

We conclude this part on the two-dimensional lattice by mentioning that, while
fully symmetric wave functions may easily describe situations with collinear
magnetic order (we showed the case of N\'eel order), it is much less trivial 
to obtain noncollinear magnetic states. Usually, noncollinear orders appears in
frustrated lattices, which break the Marshall sign rule. While these states
may be easily captured by explicitly breaking the symmetry in the variational
wave function~\cite{tocchio2013b,ghorbani2015}, we could not succeed at reproducing
them within a fully symmetric state.

\section{Conclusions}\label{sec:conclusions}

In this paper, we have shown that Jastrow-Slater wave functions, constructed
by applying a Jastrow factor to noninteracting fermionic states, represent
a very flexible tool to describe different phases of strongly correlated
systems. In particular, it is possible to capture phases with broken
symmetries even when these variational states are symmetry-invariant.
We reported two classes of examples. In the first one, which applies to 
itinerant systems (i.e., Hubbard-like models), we showed that charge or 
orbital order may naturally emerge from a short-range Jastrow factor. 
The existence of a phase transition and the stabilization of a symmetry-broken 
phase can be related to a simple mapping between quantum averages and an 
effective classical partition function, where the strength of the Jastrow 
factor is directly related to an effective classical temperature. In this 
case, the configurations that are generated along the Monte Carlo simulation 
face a breaking of ergodicity, when the Jastrow pseudo-potential is strong 
enough. This is exactly the same phenomenology of classical systems that 
undergo phase transitions at low temperatures (e.g., the two-dimensional 
Ising model).

In the second class, which applies to spin systems (i.e., Heisenberg-like
models), we have illustrated that antiferromagnetism in two dimensions and 
dimerization in one dimension may be generated in fully projected wave functions. 
In this case, there is no classical mapping to guide physical 
intuition. The emergence of a finite magnetization in two dimensions is due 
to the presence of the full Gutzwiller projector and of long-range singlets 
that create a sizable entanglement in the variational wave function, similarly 
to what happens within the bosonic LDA wave function. In contrast to the 
previous example on charge/orbital order, in the magnetic order case there 
is no broken ergodicity in the Monte Carlo single-particle moves (e.g., all 
the spin components have exactly zero expectation value). This may be ascribed 
to the fact that the magnetic order is Heisenberg-like while the charge order 
is Ising-like; the former can easily overcome the potential barrier while the 
latter cannot. Nevertheless, as for the LDA wave function, a finite magnetization 
is achieved when singlets are correlated at long distances. We would like to 
conclude by emphasizing the fact that the presence of the full Gutzwiller projector 
of Eq.~(\ref{eq:fullgutz}) is necessary to obtain magnetic order; otherwise, the 
soft Gutzwiller term~(\ref{eq:softgutz}) can only change spin-spin correlations 
at short distances, implying a vanishing magnetization in the thermodynamic 
limit.

\acknowledgments

We thank S. Sorella for interesting discussions at the early stage of this
project. L.F.T. and F.B. acknowledge the support of the Italian Ministry of 
Education, University, and Research through Grant No.\ PRIN 2010 2010LLKJBX.
R.K., R.V. and C.G acknowledge the support of the German Science Foundation (DFG) 
through Grant No.\ SFB/TRR49. The variational Monte Carlo code used for the 
magnetic calculations is based on a code first developed by D. Tahara.
We thank A. Paramekanti for having drawn to our attention 
Ref.~\onlinecite{motrunich2004}. F.B. and R.V. acknowledge the KITP program 
{\it New Phases and Emergent Phenomena in Correlated Materials with Strong 
Spin-Orbit Coupling} and the National Science Foundation (NSF) under Grant No. NSF 
PHY11-25915.

\end{document}